\documentclass[aps,pra,twocolumn,showpacs,superscriptaddress,letterpaper]{revtex4-1}

\usepackage{mathrsfs,graphicx,amssymb}

\usepackage[fleqn]{amsmath}
\usepackage{calrsfs}

\begin{document}

\title{Laser induced electron diffraction:\\ a tool for molecular orbital imaging}

\author{M. Peters}
\affiliation{Universit\'e Paris-Sud, Institut des Sciences Mol\'eculaires d'Orsay (CNRS), F-91405 Orsay, France}
\affiliation{D\'epartement de Chimie, Universit\'e Laval, Qu\'ebec, Qu\'ebec, Canada G1V 0A6}

\author{T. T. Nguyen-Dang}
\affiliation{D\'epartement de Chimie, Universit\'e Laval, Qu\'ebec, Qu\'ebec, Canada G1V 0A6}

\author{E. Charron}
\affiliation{Universit\'e Paris-Sud, Institut des Sciences Mol\'eculaires d'Orsay (CNRS), F-91405 Orsay, France}

\author{A. Keller}
\affiliation{Universit\'e Paris-Sud, Institut des Sciences Mol\'eculaires d'Orsay (CNRS), F-91405 Orsay, France}

\author{O. Atabek}
\affiliation{Universit\'e Paris-Sud, Institut des Sciences Mol\'eculaires d'Orsay (CNRS), F-91405 Orsay, France}

\date{\today}

\begin{abstract}
We explore the laser-induced ionization dynamics of N$_2$ and CO$_2$ molecules subjected to a few-cycle, linearly polarized, 800\,nm laser pulse using effective two-dimensional single active electron time-dependent quantum simulations. We show that the electron recollision process taking place after an initial tunnel ionization stage results in quantum interference patterns in the energy resolved photo-electron signals. If the molecule is initially aligned perpendicular to the field polarization, the position and relative heights of the associated fringes can be related to the molecular geometrical and orbital structure, using a simple inversion algorithm which takes into account the symmetry of the initial molecular orbital from which the ionized electron is produced. We show that it is possible to extract inter-atomic distances in the molecule from an averaged photon-electron signal with an accuracy of a few percents.
\end{abstract}

\maketitle

\section{Introduction}

The proposal and experimental realizations of new ultra-fast molecular imaging techniques based on  electron dynamics in intense ultra-short laser pulses have been the subject of  intense research activity in the last decade\,\cite{J.Mod.Opt.57.916}. Due to the ability they offer to visualize, at the atomic scale, the ultrafast molecular dynamics taking place in a chemical reaction or during an internal rearrangement process, these techniques obviously will find many applications in biological, chemical and material sciences. Conventional methods used to achieve atomic resolution, for example X-ray or electron diffraction, are much more limited in time resolution than ultrafast electron-dynamics based imaging techniques. One of those new imaging approaches consists of probing  a molecule  by its own electrons, strongly driven and ionized by an intense laser pulse\,\cite{J.Phys.B.44.ip}.

This idea originates from the so-called three-step electron rescattering mechanism\,\cite{JETP.Lett.45.404, Phys.Rev.Lett.70.1599, Phys.Rev.Lett.71.1994}. It is now widely accepted that, during strong-field atomic or molecular ionization by an optical field, a free electronic wave packet is formed each time the laser field passes its maximum value. These wave packets, initially accelerated by the field, have a large probability to return to the vicinity of the parent ion when the electric field reverses its sign, half a cycle later. This creates a series of recollision processes, taking place with very high electron kinetic energies. The consequences of these electron-ion rescattering events are diverse and can lead to several different physical processes: elastic scattering of the returning electron, electronic excitation of the parent ion, non-sequential double ionization (NSDI), but also recombination of the recolliding electron with the orbital from which it was extracted, thus producing high order harmonics (HHG).

Using a few-cycle laser pulse, the measurement of the outcome of the electron-ion rescattering events, which take place only half an optical cycle after ionization, can be seen as an ultra-fast probe of the molecule by its own electrons. It has for instance been realized that the high order harmonic spectrum carries very precise informations on the molecular structure and dynamics\,\cite{Nature.432.867}, with exceptional temporal and spatial resolutions. The harmonic spectrum is indeed related to the transition dipole between the ground and continuum electronic states. Assuming a simple form for the continuum wave functions allows one to reconstruct the molecular ground state from a complete measurement of the harmonic phases and amplitudes. This method was proposed and used for the first time in 2004 by the group of P. B. Corkum for the accurate experimental reconstruction of the highest occupied molecular orbital (HOMO) of N$_2$\,\cite{Nature.432.867}. Even though the assumptions made in such a reconstruction method are still a matter of debate, this experimental breakthrough has generated a wealth of promising new theoretical and experimental studies\cite{Phys.Rev.A.66.023805, Nature.435.470, Phys.Rev.Lett.95.153902, Science.322.1232, Nature.460.972, Nat.Phys.6.200}.

For instance, the intensity modulation of the harmonic spectra generated by aligned linear molecules was shown to be related to a multi-center molecular interference which can be used to measure the internuclear separation in the molecule\,\cite{Phys.Rev.A.66.023805, Nature.435.470, Phys.Rev.Lett.95.153902}. It should however be mentioned that the measured harmonic spectrum may be rather complex since it carries informations not only on the HOMO orbital, but also on lower lying orbitals (\mbox{HOMO-$n$}, $n=1,2,\ldots$), which, depending on the molecule itself and on its orientation, may contribute significantly or not to the HHG spectrum\,\cite{Science.322.1232, Nature.460.972, Nat.Phys.6.200}.

Elastically scattered electrons also carry  structural informations on  the molecule, since their momentum distribution can be influenced by interferences between the amplitudes arising from the different scattering centers constituted by the nuclei. This imaging technique, first proposed in 1996 by T. Zuo, A. D. Bandrauk and P. B. Corkum, was called ``\textit{Laser-induced electron diffraction}'' (LIED)\,\cite{Chem.Phys.Lett.259.313}. This proposal was investigated theoretically by different groups\,\cite{Phys.Rev.A.66.051404, J.Phys.B.37.L243, Phys.Rev.Lett.93.223003, Phys.Rev.Lett.94.073004, Phys.Rev.A.82.051404}. It was first shown in restricted two-dimensional quantum simulations that clear signatures of two-center interferences could be distinguished in the re-scattered electron angular distributions\,\cite{Phys.Rev.A.66.051404}. Other investigations analyzed how these interferences could be used to image the structure of small molecules\,\cite{J.Phys.B.37.L243}. This question was then addressed in full three-dimensional quantum simulations, and a relatively simple analytic transformations was proposed to extract the internuclear separation of a diatomic molecule from the full three-dimensional electron momentum distribution\,\cite{Phys.Rev.Lett.93.223003, Phys.Rev.Lett.94.073004}. The first successful experimental demonstration of LIED was finally reported in 2008 from aligned O$_2$ and N$_2$ diatomic gases\,\cite{Science.320.1478}.

Contrarily to conventional electron diffraction, the ability to extract structural information from the outcome of a rescattering event, be it harmonic generation or laser-induced electron diffraction, relies heavily on the ability to align the molecule in the laboratory frame.  
 Assuming a realistic molecular alignment, we have recently demonstrated an accurate, simple and robust method to extract the molecular structure from the photo-electron spectra of a laser-driven linear, symmetric, polyatomic molecule\,\cite{Phys.Rev.A.83.051403}. We have also shown that the detailed structure of the diffraction image reflects the symmetry of the molecular orbital from which the recolliding electron emanates. 
 
 The present paper gives a more detailed theoretical analysis of this effect, illustrated first on the diatomic N$_2$ molecule then on the triatomic CO$_2$ molecule, one of the simplest linear polyatomic system. It is organized as follows. The physical process of laser-induced electron diffraction is described in Sec. \ref{sec2}. Our two-dimensional effective quantum model and the optimization of the model parameters are presented in Sec. \ref{sec3}. This section also presents  results of our simulations on N$_2$, taken as an illustrative example of a model diatomic molecule, together with a number of assumptions leading to a simplified treatment of the electronic diffraction dynamics. In Sec. \ref{sec4}, we present our theoretical predictions and analysis of the electron momentum distribution for the triatomic molecule CO$_2$. The spectra obtained with the three highest molecular orbitals are successively discussed, and the influence of various model assumptions are then presented. We conclude the work with a summary of our findings in section \ref{sec5}.

\section{LIED and the three-step mechanism}
\label{sec2}

\begin{figure}[!t]
\begin{center}
\includegraphics[width=0.99\columnwidth]{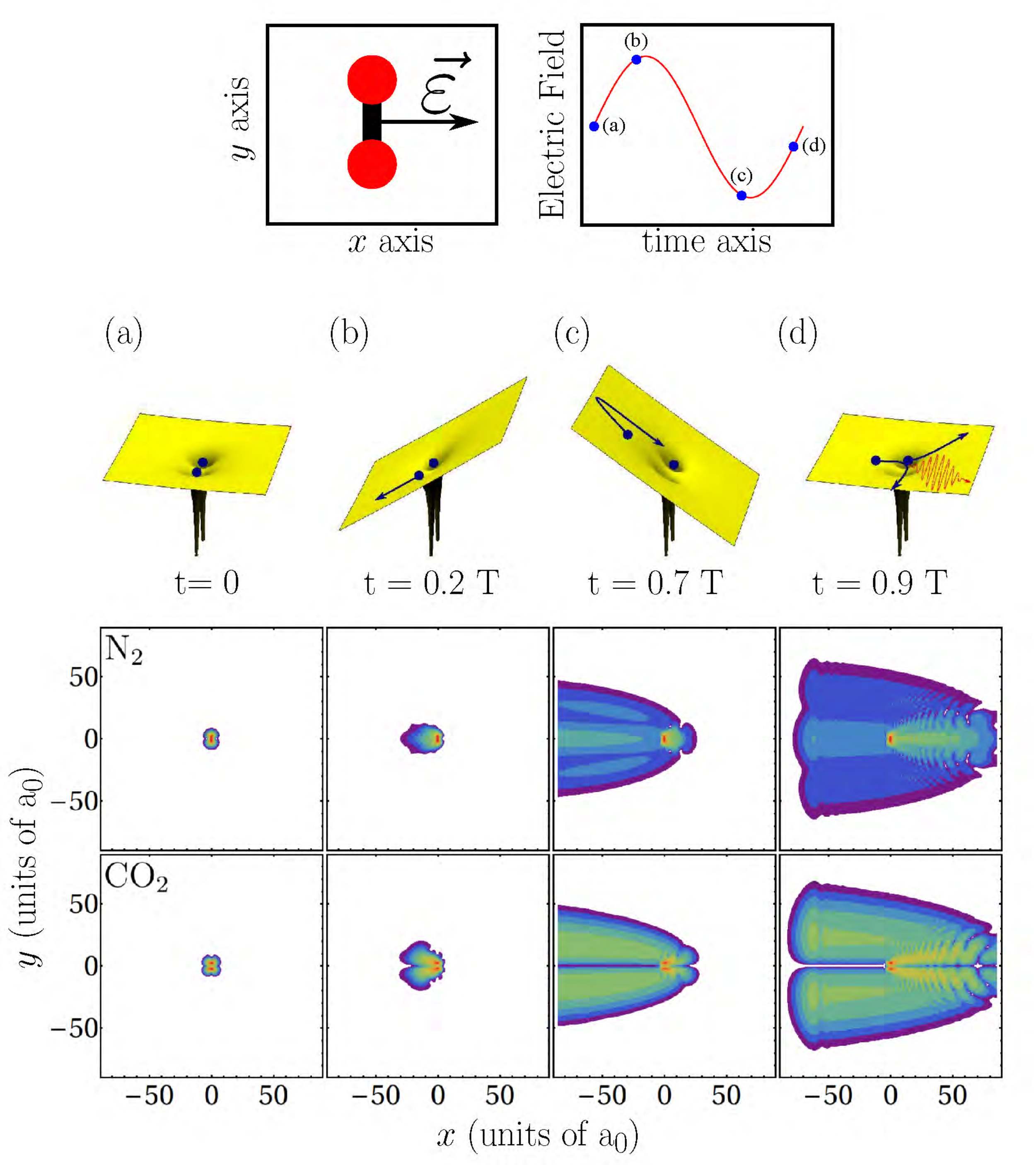}
\caption{\label{fig1}(Color online) Schematic view of the three-step mechanism as inspired from \cite{CKNP2007}. The upper part displays the positioning of the molecule and the electric field polarization vector in the laboratory frame, together with the time variation of its amplitude ($\omega_L = 0.06$\,a.u., $E = 0.15$\,a.u.). Panels (a) to (d) illustrate the time evolution of the field-dressed two-center Coulomb potentials corresponding to different snapshots within the optical cycle, as indicated by the dots on the field evolution. Numerical results for the square modulus of the electron wave packet are shown on the bottom part, starting initially from the HOMO of N$_2$ (upper part) and of CO$_2$ (lower part). See text for details.}
\end{center}
\end{figure}

We start with a brief  review of the so-called three-step mechanism, the principal idea from which the above dynamical molecular imaging schemes originate. Figure\,\ref{fig1}, inspired from the work of P. B. Corkum and F. Krauss\,\cite{CKNP2007}, depicts the electron motion occurring in the combined oscillating force of an external laser field and that deriving from its binding potential, the latter being chosen as the double-well Coulomb potential of  a laser driven  diatomic molecule, N$_2$ for instance. 

The coordinate system $(x,y)$ refers to the plane defined by the molecular axis and the polarization vector $\vec\varepsilon$ of the linearly polarized electric field. The molecule is assumed to be initially aligned along the laboratory $y$-axis, while the intense laser pulse, which drives the electron dynamics, is polarized along the laboratory $x$-axis, as illustrated in the first row of Fig.\,\ref{fig1}. Due to the form of the radiative interaction, the subsequent electronic dynamics takes place essentially in the $(x,y)$-plane. Also illustrated in the first row of Fig.\,\ref{fig1} is a single-cycle laser pulse defined as $\vec{\mathscr{E}}(t) = E\sin(\omega_L t)\,\vec\varepsilon$ for $t \leqslant T$, where $T = 2 \pi/\omega_L$ is the period of the field. Here, $\omega_L=0.06$\,a.u., corresponding to the wavelength $\lambda = 800$\,nm of a Ti-Sapphire laser, and $E=0.15$\,a.u., corresponding to a  peak intensity $I= 8 \times 10^{14}$\, W/cm$^2$. The middle panels of Fig.\,\ref{fig1} display, for specific times within the optical cycle, the total double-well Coulomb potential distorted by the radiative interaction potential $\vec{r} \cdot \vec{\mathscr{E}}(t)$, written in the length gauge, $\vec r$ being an electron coordinate. In addition to the potential surfaces, the electron motion is schematically indicated by classical trajectory-style representations. Panels (a) to (d) refer to the different times  indicated by the labeled dots placed on the electric-field waveform on the first row of the figure.

Panel (a) is for $t = 0$ and corresponds to the field-free Coulomb potential. Panel  (b) is for $t = 0.2 \ T$,  a time close to the field peak intensity. The total Coulomb plus field interaction potential is then characterized by a maximum positive slope. The molecule is ionized around that time, when the electric field amplitude is sufficiently high to lower the potential barrier allowing tunnel ionization. The positive electric field then accelerates the electron in the  ``forward'' direction, $x < 0$. At $t = T/2$, the electric field changes its sign and so does the force on the electron, which is accelerated from then on in the opposite direction $ x > 0$. The potential slope is then negative and the electron trajectory is classically accelerated back to the double-well region. The wave packet still extends in the forward direction until some time later,  $t = 0.7 \ T$,  panel (c) when the absolute value of the slope of the potential reaches its maximum. This is the classical turning point of the electron trajectory. Panel (d), for $t=0.9 \ T$, illustrates the subsequent recollision process of the electron with the nuclei. The trajectories starting around $t = T/4 + T/20$ lead to the most energetic recollision events\,\cite{J.Phys.B.44.ip}. The system has thus produced its own electron gun. In addition to this, at times close to $3T/4$, the field is again large enough to significantly lower the potential energy barrier through which tunnel ionization could occur. There is then a possibility for a second electronic wave packet to be launched and accelerated in the backward direction. Finally, it is important to note the asymmetry of the forward and backward scattering. The electron wave packet launched at a time close to $T/4$ is accelerated in the forward direction from the time of its birth and until it starts to return to the core. This oscillating wave packet interferes with a second wave packet produced around $t = 3T/4$. This explains the observation of interference patterns in the backward direction $(x > 0)$ in the lower rows of Fig.\,\ref{fig1}.

\begin{figure}[ht]
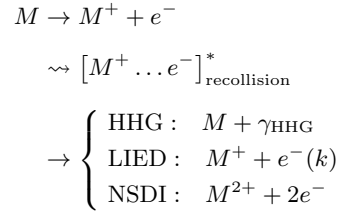

\begin{eqnarray}
 M & \rightarrow      & M^+ + e^- \nonumber\\[0.2cm]
   & \rightsquigarrow & \left[M^+ \dots e^-\right]^*_{\mathrm{recollision}} \nonumber\\[0.2cm]
   & \rightarrow      & \left\{
                        \begin{array}{l}
                        \mathrm{HHG:}  \quad M + \gamma_{\mathrm{HHG}}\\[0.1cm]
                        \mathrm{LIED:} \quad M^+ + e^-(k)\\[0.1cm]
                        \mathrm{NSDI:} \quad M^{2+} + 2e^-
                        \end{array}
                        \right.\nonumber
\end{eqnarray}
\caption{\label{fig2}Three-step electron rescattering mechanism.}
\end{figure}

A wealth of complex phenomena are induced by the recollision process in this three-step mechanism\,\cite{JETP.Lett.45.404, Phys.Rev.Lett.70.1599, Phys.Rev.Lett.71.1994} depicted in Fig.\,\ref{fig2}. Indeed, the recollision may result - through the temporary recapture of the electron - in the formation of a transient excited species $\left[M^+ \dots e^-\right]^*$ leading to several possible decay channels. One of them corresponds to the definite recapture of the electron with the emission of high order harmonics (HHG) currently exploited in the generation  of attosecond pulses\,\cite{CKNP2007}. Another possible process is non-sequential double-ionization (NSDI)\,\cite{WPRL2000, MPRL2000}. Yet another possibility, which is the subject of this paper is laser induced electron diffraction (LIED)\,\cite{Science.320.1478}, yielding a re-scattered electron with high momentum  $k$. It has also been observed that the recollision process may even leave the electron in highly excited Rydberg states, as the ionized electron, being decelerated over many laser cycles, may be recaptured when the field is over. This last process has been called ``\textit{frustrated tunnel ionization}''\,\cite{MPRL2009,EarXiv2011}.

In the following, we wish to demonstrate how photoelectron momentum distributions of a laser-driven (and ionized) linear molecule reflect the geometrical and orbital structures of the molecule. The aim is to show how LIED can be used as a nuclear geometry imaging technique, and to establish a reliable procedure to retrieve geometrical informations from the photoelectron spectrum. We will refer to this procedure as an inversion algorithm.

\section{Theoretical model}
\label{sec3}

The laser-driven electron dynamics and the photoelectron spectra are calculated using a number of simplifying assumptions. These fall roughly into two classes: those pertaining to the model of the physical system itself, and those relevant to questions of experimental feasibility.

The two assumptions that may be of most experimental concern are the hypothesis of a perfect molecular alignment and that of a few cycle pulse excitation. As will be shown later, the default to alignment is a major limitation for the quality requirements of the imaging technique. Indeed, different alignment angles with respect to the polarization vector lead to different diffraction patterns, and we will have to assess the robustness of the results of our LIED analysis with respect to misalignment of the molecule.  Another experimental challenge is the production of few-cycle laser pulses. We started out with observations made with a single-cycle pulse, for which the LIED is most transparent. The inclusion of several cycles will induce  complicated diffraction patterns due to several back and forth   oscillations of the electronic wave packet along  with multiple recollision events. One can expect, in a near future, more sophisticated molecular alignment techniques, on one hand, and the production of intense near single-cycle pulses, on the other hand. How the contrast of LIED  images is affected by the lack of perfect alignment and the effect of multi-cycle pulses on the diffraction spectra will however be discussed in detail in the second part of this article and illustrated  explicitly for the case of the CO$_2$ molecule.

As to the model used for the physical (molecular) systems, the major limitation is the restricted number of degrees of freedom for the description at two levels: First, concerning nuclear degrees of freedom, a fixed nuclei approximation is used and the nuclear vibrational and rotational dynamics are completely neglected. This is justified by the ultrafast time scale of the electronic LIED motion (attosecond range) as compared to the vibrational and rotational time scales (femtosecond to picosecond range).  Second,   a single  active electron (SAE) model with a soft-Coulomb pseudo-potential is used to describe the electron dynamics, and the associated time-dependent Schr\"odinger equation (TDSE) is solved  in two-dimensions.

\subsection{SAE Hamiltonian}

Electronic wave packets are generated from an initial wave function $\Phi(\vec r , t_0)$ by writing
\begin{equation}
\Phi(\vec r , t) = U_H(t, t_0) \Phi(\vec r , t_0)\,, \label{Phi}
\end{equation}
$U_H$ being the unitary time evolution operator satisfying the TDSE
\begin{equation}
\dot\imath\frac{\partial}{\partial t} U_H(t, t_0) = H(\vec r, t) U_H(t, t_0), \qquad U_H(t_0, t_0)=1
\end{equation}
with a SAE-Hamiltonian of the form
\begin{equation}
\label{Hamiltonian}
H(\vec r, t) = \frac{p^{2}}{2} + \sum_{\alpha = 1}^{3} \frac{- Z_{\alpha}(\vec r_{\alpha})}
               {\sqrt{|\vec r_{\alpha}|^2 + a_{\alpha}^2}} + \vec r \cdot \vec{\mathscr{E}}(t)\,,
\end{equation}
$\vec p$ being the electron momentum operator. Atomic units are used throughout the paper unless stated otherwise. $\alpha=1,...,N$ labels the nuclei at fixed positions $\vec{\rho}_{\alpha}$ and $\vec r_{\alpha}=\vec r- \vec \rho_{\alpha}$ where $\vec r \equiv (x,y)$ denotes the electron position in the two-dimensional $xy$-plane. The position-dependent effective charge $Z_{\alpha}(\vec r)$ is chosen of the following analytical form
\begin{equation}
\label{eff_charge}
Z_{\alpha}(\vec r) = Z_{\alpha}^{\infty} + \left( Z_{\alpha}^{0} - Z_{\alpha}^{\infty} \right)
                     \exp \left[-\frac{|\vec r - \vec\rho_{\alpha}|^2}{\sigma_{\alpha}^2} \right]\,,
\end{equation}
where $Z_{\alpha}^{\infty}$ denotes the effective nuclear charge of the nucleus $\alpha$ as seen by an electron at infinite distance. $Z_{\alpha}^{0}$ is the bare charge  of nucleus $\alpha$ and $\sigma_{\alpha}$ is a parameter which characterizes the decrease of the effective charge of that nucleus   with  distance, and is introduced to account for distance-dependent electron-electron screening effects. The value of $Z_{\alpha}^{\infty}$ is derived from a Mulliken analysis carried out in an \emph{ab-initio} study performed on the parent ion. As a result, the sum $\sum_{\alpha} Z_{\alpha}^{\infty}$, denoting the total charge of the parent molecular ion, is equal to 1.

The SAE potential defined in Eqs.\,(\ref{Hamiltonian}) and (\ref{eff_charge}) thus represents the force field seen by the active electron. It takes into account the interaction with the different nuclei of the molecule and with the $N-1$ other electrons, which screen the nuclei. Table\,\ref{param} summarizes the values of all parameters defining the SAE potential for the case of the N$_2$ and CO$_2$ molecules.

\begin{table}[!h]
\begin{center}
\begin{tabular}{|c||c|c||c|c|c|}
\hline
Molecule               & \multicolumn{2}{c||}{N$_2$} & \multicolumn{3}{c|}{CO$_2$}\\
\hline
Atom                   & N       & N       & O       & C       & O\\
\hline\hline
$a_\alpha$ (a.u.)      & 1.2     & 1.2     & 1.0     & 1.0     & 1.0\\
$\sigma_\alpha$ (a.u.) & ~0.700~ & ~0.700~ & ~0.577~ & ~0.750~ & ~0.577~\\
$Z^0_\alpha$           & 7       & 7       & 8       & 6       & 8\\
$Z^{\infty}_\alpha$    & ~0.500~ & ~0.500~ & ~0.173~ & ~0.654~ & ~0.173~\\
\hline
\end{tabular}
\end{center}
\caption{\label{param}Values of all parameters used for the single active electron effective potentials of N$_2$ and CO$_2$.}
\end{table}

\subsection{Electron wave packets}

\subsubsection{Wave packet propagation algorithm}

The third-order split-operator technique \cite{splitop} is used to solve the TDSE together with a Volkov-type asymptotic analysis \cite{volkov}. 
Following Refs.\,\cite{cutWF} and\,\cite{Keller}, one   divides the electronic space into two regions, an asymptotic region $(A)$ where the Coulomb potential energy is neglected  and an internal region $(I)$ where all interaction potentials act. The wave function (\ref{Phi}) is split accordingly as follows
\begin{equation}
 \Phi(x, y , t) = \Phi_I(x, y , t) + \Phi_A(x, y , t)\,,
\end{equation}
where $\Phi_I(x, y , t)$ is non-zero in the internal region only and $\Phi_A(x, y , t)$ is non-zero in the asymptotic region only. This is done through some smooth function $f(x,y)$ which is equal to 1 in the interaction region and equal to 0 in the asymptotic one, such that
\begin{subequations}
\label{PhiIA}
\begin{eqnarray}
 \Phi_I(x, y , t) & = & f(x,y) \times \Phi(x, y , t)\\
 \Phi_A(x, y , t) & = & \left[1-f(x,y)\right] \times \Phi(x, y , t)\,.
\end{eqnarray}
\end{subequations}
The linearity of the Schr\"odinger equation allows one to propagate the wave functions $\Phi_I(x, y , t)$ and $\Phi_A(x, y , t)$ separately, using  different methods as suited to each region. In the internal region, the electronic wave function $\Phi(x, y , t)$ is then replaced by its internal part $\Phi_I(x, y , t)$, and its propagation is done using the third-order split-operator algorithm. Any portion of this propagated function that acquires a significant amplitude in the asymptotic region is collected as the asymptotic component $\Phi_A(x,y,t)$. Each newly removed asymptotic component is then propagated separately using the analytical Volkov solution\cite{Keller} and added to the previously cut-out and propagated asymptotic components. The final time of propagation is chosen such that the accumulation of the outgoing components is converged. Typically, the temporal propagation lasts a few tens of femtoseconds after the end of the pulse. As for the function $f(x,y)$,  an infinite number of choices exists, with only one limitation: that its spatial variation must be slow enough to avoid unphysical quantum reflexions but fast enough to separate efficiently the internal region from the asymptotic domain\,\cite{Keller}.

\subsubsection{Initial state}

\begin{figure}[!t]
\begin{center}
\includegraphics[width=0.99\columnwidth]{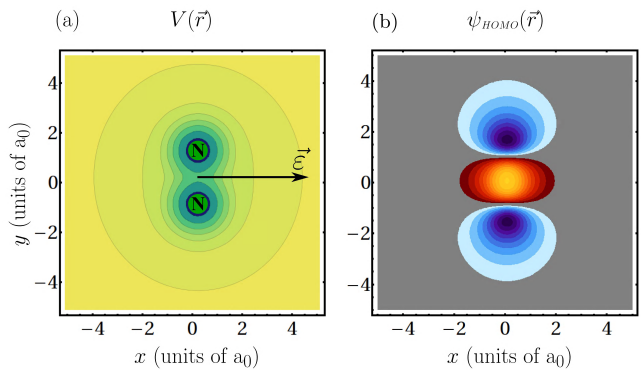}
\caption{\label{fig3}(Color online) (a) Soft-Coulomb potential associated with N$_2$. (b) Two-dimensional representation of the HOMO probability amplitude distribution of N$_2$. The change of color denotes a change of sign of the orbital wave function.}
\end{center}
\end{figure}

The initial state $\Phi(x, y , 0)$ of Eq.(\ref{Phi}) is typically taken as one of the HOMO-$n$ $(n=0,1,2)$ orbitals of the molecule. In the two-dimensional SAE model with a soft-Coulomb pseudo-potential,  as shown in Fig.\,\ref{fig3}(a) for N$_{2}$, this initial state is obtained by integrating the time-dependent Schr\"odinger equation in imaginary time $( t \rightarrow -it )$ using the split-operator technique \cite{splitop}. Starting with an arbitrary trial initial wave function of appropriate symmetry and requiring this wave function to remain normalized, the propagated wave function will converge to the lowest energy state of the same symmetry, the excited components dying off exponentially \cite{Davies1980, Lehtovaara2007}. Excited states of this symmetry can also be obtained by propagating different initial trial wave functions of the correct symmetry with an additional Gram-Schmidt orthogonalization to lower energy states after each time step.

In the case of N$_2$, only the HOMO orbital is considered as the initial state. In a complete $N=14$-electron description, it is the third orbital, (in order of increasing energy), of $\sigma_g$ symmetry, and is noted $3\sigma_g$. This HOMO orbital  is characterized by a symmetry of revolution about the molecular axis and by two nodal planes orthogonal to this  axis. The HOMO calculated, by imaginary-time wave packet propagation, in this SAE 2D model, has indeed two nodes  on the $y$-axis, as shown in Fig.\,\ref{fig3}(b). The energy of this  HOMO orbital depends on what is chosen for the parameters of the soft-Coulomb potential. Here, the value of $\sigma_{\alpha}$ is optimized in such a way that the energy of the HOMO approaches the experimental ionization potential of this particular orbital, determined by electron impact \cite{NIST} while all other parameters remain fixed. For the case of N$_2$, since only one molecular orbital will be considered, the exact energy value could be obtained.

In the case of the CO$_2$ molecule, three different molecular orbitals, the HOMO, HOMO-$n$, with $n=1,2$, were considered separately as possible initial states of the SAE dynamics. The HOMO, of $\pi_g$ symmetry, is doubly degenerate in 3D, and, in the crudest LCAO scheme, can be expressed as the antisymmetric combination of the Oxygens' $2p$ orbitals that are oriented perpendicular to the molecular axis. The HOMO is   characterized by two nodal planes: one containing the  molecular axis and another one perpendicular to and bisecting the molecular axis. The  HOMO-1 is of $\pi_u$ symmetry, and is also doubly degenerate. It can be expressed as the  symmetric linear combination of the Oxygen and Carbon $2p$ orbitals that are perpendicular to the molecular axis (for example the $2p_x$ atomic orbitals). These orbitals are characterized by a nodal plane containing the molecular axis.
The HOMO-2, of $\sigma_{u}$ symmetry, is non degenerate and can be expressed as a  linear combination of the Carbon $2p$ orbital  oriented along the direction of the molecular axis with an anti-symmetric combination of the Oxygens' $2s$ and a symmetric one of the Oxygens' $2p$ orbitals oriented along the molecular axis.   It has a nodal plane perpendicular to and bisecting the molecular axis.

Enforcement of these symmetry properties, together with imaginary-time propagation, yield the three desired molecular orbitals depicted in Fig.\,\ref{fig5}. Table \ref{param} lists the potential parameters optimized for this case. This optimization of the soft-Coulomb potential's parameters was done  such that the ionization potentials calculated with  it agree satisfactorily and simultaneously for all three orbitals with the experimental measurement \cite{NIST}. Therefore, with the optimal values  given in Table \ref{param}, one finds 13.6\,eV, 17.8\,eV and 22.3\,eV for the ionization potentials of the HOMO, HOMO-1 and HOMO-2 respectively, as compared to their respective measurement of 13.8\,eV, 19.7\,eV and 20.3\,eV, as reported in \cite{NIST}. The maximum relative deviation from the experimental data remains within 10\% in this case. In principle, the effective potential seen by the active electron depends on the molecular orbital under scrutiny, as in Hartree-Fock theory, where exchange and Coulomb parts of the Fock operator depend on the unknown orbitals. The simpler picture adopted here is thus expected to have some limitations in the simultaneous description of many orbital states \cite{Saugout2008}. For the purpose of this work, where the feasibility of LIED as a molecular imaging scheme is assessed,  this description must be sufficiently accurate to capture the essentials of the electron dynamics. 

\subsubsection{Spatial wave packet: Diffraction fringe structures}

Using the propagation algorithm described above, electronic wave packets  were calculated (for both N$_2$ and CO$_2$) on a spatial grid of 800 points on each axis, spanning $-170\,a_0$ to $170\,a_0$. The split-operator algorithm for the time-evolution operator uses the time interval $\delta t = 0.05$\,a.u. Snapshots of the wave packet obtained at the  different times of a single-cycle pulse as indicated on the waveform of the field shown on the first row of Fig.\,\ref{fig1} are given in the lower panels of this figure, for N$_2$ and CO$_2$.  The series of spatial wave packet contour plots  does reflect the sequence of classical events that was discussed in detail in section \ref{sec2}, corresponding to different phases of  the three-step mechanism. Thus
at time $t \simeq 0.2\,T$, the electronic wave packet starts its forward motion (toward $x<0$), while at $t=0.7\,T$  it reached its maximum extension   in the forward $(x<0)$ direction. Of note is that at  $t=T$   an interference pattern is  observed between the direct and rescattered components of the ionized wave packet. 

The most important observations to be drawn from these snapshots describing the LIED process are: ($i$) the fringes obtained in the forward direction [$x<0$ in the third and fourth rows of Fig.\,\ref{fig1}(c)] as a signature of the lobes of the corresponding HOMO orbitals acting as electron ejection sources, on one hand, and ($ii$) the rich pattern of interference fringes in the backward direction [$x>0$ in the third and fourth rows of Fig.\,\ref{fig1}(d)] resulting from the electron diffraction analogous to Young's slit experiments, on the other hand. In this analogy, the slits are replaced by atomic diffusion centers. In the case of N$_2$, the HOMO shows three lobes along $y$-axis leading to three main fringes in the diffraction pattern (see Fig.\,\ref{fig1}, third row, right panel). On the contrary, the HOMO of CO$_2$ displays only two lobes along the $y$-axis and its diffraction pattern shows two families of fringes with a nodal structure along $y=0$ (see Fig.\,\ref{fig1}, fourth row, right panel).

\subsection{Electron momentum distribution and its qualitative analysis}

\begin{figure}[!t]
\begin{center}
\includegraphics[width=0.99\columnwidth]{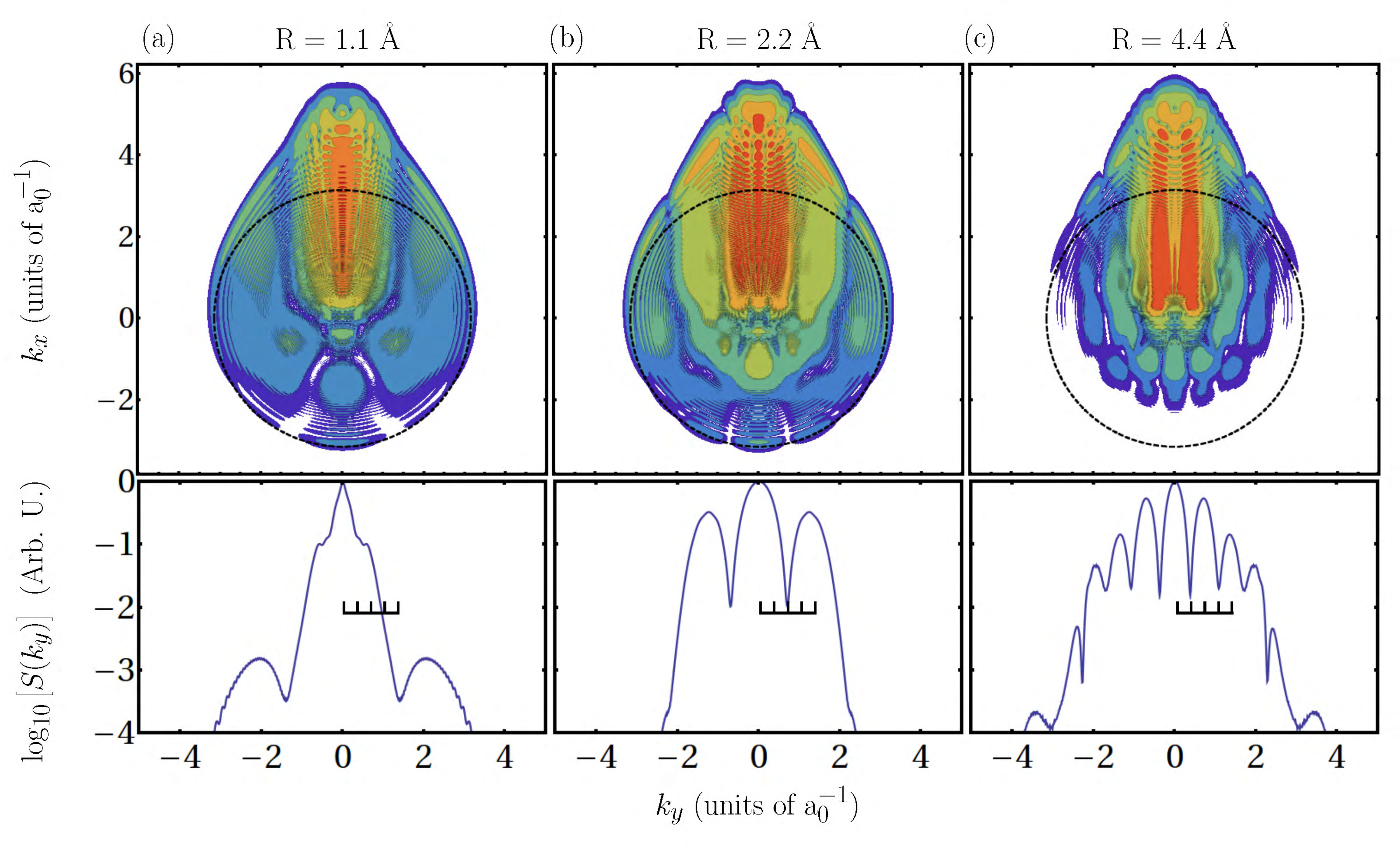}
\caption{\label{fig4}(Color online) Electron diffraction patterns $|\tilde{\Phi}_A(k_x,k_y)|^2$ for N$_2$ in the momentum space representation (upper panels) and for three different inter-atomic distances $R$ from equilibrium to stretched geometries and under the same excitation conditions as Fig.\,\ref{fig1}. Dashed circles indicate points with final energy equal to the maximum classical recollision energy. The lower panels display the averaged diffraction spectra $S(k_y)$ on a logarithmic scale normalized to the maximal value. Tick mark rulers are given to facilitate the counting of peaks within a given $k_y$-interval.}
\end{center}
\end{figure}

The experimentally accessible observable, through time-of-flight electron velocity mapping is the diffraction pattern in the reciprocal momentum space ($k_x$, $k_y$). Theoretically, this is observed in the electronic asymptotic momentum distribution,  defined as the Fourier transform $\tilde{\Phi}_A(k_x,k_y)$ of the asymptotic electron wave packet $\Phi_A(x,y,t_f)$ at the end of the propagation $(t=t_f)$
\begin{equation}
 \tilde{\Phi}_A(k_x,k_y) \propto \int \Phi_A(x,y,t_f) e^{-ik_xx}e^{-ik_yy} dx dy
\end{equation}
In the following, the tilde ($\tilde{~}$) symbol will denote a Fourier transform. In practice, the final time $t_f$ needs not be infinite, but is taken sufficiently large to ensure convergence. The upper panels of Fig.\,\ref{fig4} display the square modulus of $\tilde{\Phi}_A(k_x,k_y)$, for three different internuclear distances $R$ of N$_2$: (a) for the molecule in its equilibrium geometry $R=1.1$\,\AA\ ; (b) and (c) for a stretched N$_2$ with $R=2.2$\,\AA\ and $4.4$\,\AA, respectively.

It was shown that the overall three-step recollision process leads to a maximum collision energy given as $3.17\,U_p$ \cite{Krause1992, Kulander1993, LHuillier1993}, where $U_p = E^2/4\omega_L^2$ is the electron ponderomotive energy. Assuming elastic collisions distributed over all angles of the $(x,y)$-plane, this defines a circle ${\cal C}(k_x,k_y)$ in the reciprocal $(k_x,k_y)$-plane, with a well defined finite radius, such that all classical trajectories with an energy smaller than or equal to $3.17\,U_p$ remain within the boundaries of this circle. The ponderomotive motion circle associated to the maximum recollision energy for the Ti-Sapphire laser of peak intensity $8 \times 10^{14}$\, W/cm$^2$ is drawn on all the upper panels of Fig.\,\ref{fig4}.

Two observations are in order: (\textit{i}) The asymptotic electron momentum distribution basically fill up the ponderomotive motion circle, but lead also to the observation of asymmetrical high energy spectral components in the backward ($k_x>0$) scattering region going beyond a classical trajectory-type interpretation frame. This region of the spectra is filled by constructive and destructive interference patterns between backward scattering and direct ionization in the $x<0$ direction. In addition, the larger  the internuclear separation, the more pronounced  this asymmetric feature will be. (\textit{ii}) Even more important is the observation of fringes in the momentum distribution, similar to what is observed in the spatial wave packet at the end of the pulse. These fringes are   markedly well resolved in the forward ($k_x<0$) direction. Moreover the number of fringes increases with the internuclear distance. As a more quantitative analysis will make it clear, this last information is precisely the key point of the LIED imaging technique for geometrical structure determination.

Such an analysis is conducted by first averaging over $k_x$ the momentum distribution to give
\begin{equation}
 S(k_y) = \int |\tilde{\Phi}_A(k_x,k_y)|^2 \ dk_x \label{eq:average}
\end{equation}
The corresponding spectra, a diffraction pattern,  are displayed, on a logarithmic scale, on the lower panels of Fig.\,\ref{fig4}. One can relate the number of fringes observed in this diffraction pattern $S(k_y)$ with the internuclear distance: using the four tick mark rulers displayed in the lower panels of Fig.\,\ref{fig4}, it turns out that, within a typical interval of $0<k_y<1.5$\,a$_0^{-1}$, there are twice as many fringes for $R=2.2$\,\AA\ as there are, (in the same $k_y$ interval)  for $R=1.1$\,\AA, and again twice as many fringes for $R=4.4$\,\AA\ as there are for $R=2.2$\,\AA. This important observation can be summarized as
\begin{equation}
 R \propto 1 / \Delta k_y
\end{equation}
where $\Delta k_y$ denotes the momentum separation between the fringes. A complete quantitative analysis requires the determination of the proportionality factor. This will be done in the next Section.

It is however worthwhile to note that, even at this level of analysis, the diffraction spectra $S(k_y)$ may already serve as a tool for molecular imaging once the averaged diffraction spectrum has been measured at the known equilibrium internuclear distance. A possible limitation of this technique seems to be the small numbers of fringes in the spectra, which may render more difficult the measurement of the geometrical structure for short internuclear separations. This limitation can however be overcome by changing the laser parameters, such as to increase the ponderomotive energy $U_p$ and consequently the radius of the ponderomotive motion circle ${\cal C}(k_x,k_y)$ to encircle a larger area covered by recolliding trajectories. For a given frequency, the issue is to increase the peak intensity. As an indication, the laser parameters used  in obtaining the spectra of Fig.\,\ref{fig4} fulfill the requirement for the appearance of at least two fringes in ${\cal C}(k_x,k_y)$.

\section{LIED analysis and robustness}
\label{sec4}

We now give a complete proof of the feasibility of imaging  molecular geometrical and orbital structures of linear molecules. We shall first take a closer look at the diffraction pattern defined by the $k_x$-averaged asymptotic electron distribution $S(k_y)$, Eq.(8), and illustrated in Fig.\,\ref{fig4} for the case of the N$_2$ molecule.  We shall demonstrate how the geometrical and orbital structures of the molecule can be inferred from a comprehensive reading of $S(k_y)$, in what constitutes to us an  inversion algorithm for LIED. Emphasis will then be placed on results obtained for the CO$_2$ molecule, in the simple yet relatively realistic model (for the purpose of the present paper) described in the preceding section. This molecule constitutes an ideal example of a polyatomic linear symmetric molecule. The choice of this system is motivated by different considerations. With three nuclei it has enough internal degrees of freedom to make its molecular structure determination challenging. It remains however simple enough when referring to its alignment properties, due to its linear and symmetric equilibrium geometry. We will then explore and discuss the robustness of the LIED analysis when three main approximations are relaxed, at least partially. These are, in order of conceptual importance, the SAE, the perfect alignment and the single-cycle pulse excitation assumptions.

\subsection{Inversion algorithm}

The LIED inversion algorithm used here is basically related to structural properties of the initial state that are well conserved during the electron wave packet propagation, \textit{i.e.} during the time-evolution of the system under the combined effect of the Coulomb forces and of the laser field polarized perpendicularly to the molecule.

For centrosymmetric linear molecule, such as CO$_2$ and N$_2$, the addition of the time-dependent interaction with the linearly polarized laser electric field lowers the symmetry of the molecular force field from $D_{\infty h}$ to $C_{2v}$, where the $C_2$ symmetry axis corresponds to the direction $x$ of the field polarization. The symmetry operators of this subgroup ($E$, $C_2$, $\sigma_v$ and $\sigma_v'$) commute with the Hamiltonian $H(\vec r,t)$ and with the associated time-evolution operator $U_{H}$. As a consequence, if the initial state is one of the eigenstates of the field-free Hamiltonian, which necessarily transform under the various symmetry operations according to one of the four irreducible representations of the  $C_{2v}$ group, the time-dependent state that evolves from this will also transform accordingly. It is thus relevant to look at symmetry properties of the initial states considered here  for the N$_2$ and CO$_2$ molecules.

\subsubsection{{\rm N}$_2$ case}

For N$_2$, the HOMO, a $\sigma_g$ orbital in $D_{\infty h}$,  is of symmetry $a_1$ in $C_{2v}$. Its nodal structure, as seen in Fig.\,\ref{fig3}, is that of a symmetry-adapted (SA) LCAO form
\begin{equation}
\Phi_{3\sigma_g}(\vec r) \propto f_0(\vec r- \vec R/2) +  [\mathcal{S}f_0] (\vec r+ \vec R/2) \label{N2_HOMO_LCAO}
\end{equation}
where $f_0 = a\,\phi_{2s} + b\,\phi_{2p} + \ldots$ denotes a Nitrogen hybrid atomic orbital and $\mathcal{S}$ a symmetry operation of $C_{2v}$ which exchanges the two N atoms.

At any time $t$, the wave function can be expressed similarly, as
\begin{equation}
\Phi_{3\sigma_g}(\vec r,t) \propto f_t(\vec r- \vec R/2) + [\mathcal{S}f_t](\vec r+ \vec R/2)
\end{equation}
From this, we infer that the electron momentum wave function is
\begin{equation}
\tilde{\Phi}_{3\sigma_g}(\vec k,t) \propto e^{+i\vec k.\vec R/2}\,\tilde{f}_t(\vec k) + e^{-i\vec k.\vec R/2}\,[\mathcal{S}\tilde{f}_t](\vec k) \label{N2_HOMO_LCAO_t}
\end{equation}
$f_t$ can also be written as a sum of $a$- and $b$-symmetry functions as
\begin{equation}
f_t= f_t^a + f_t^b \qquad \mathrm{and} \qquad \mathcal{S} f_t= f_t^a - f_t^b
\end{equation}
and then
\begin{equation}\label{N2_HOMO_as_k_dist_2}
\tilde \Phi_{3\sigma_g}(\vec k,t) \propto \cos(k_yR/2)\tilde f_t^a(\vec k) + i \sin(k_yR/2) \tilde f_t^b(\vec k)
\end{equation}
The modulus square of this function, integrated over $k_x$, yields an oscillatory pattern that depends on the widths of the functions $\tilde f_t^a,\ \tilde f_t^b$ and on the interplay between the cosine and sine terms, which have a common period of
\begin{equation}
\Delta k_y= 4\pi/R,
\end{equation}
Without delving into the details of the actual variations of $S(k_y)$ as defined by this, we can at least infer from the period of the (modulated) oscillations in its fringe pattern the value of the internuclear distance $R$, as announced in Eq.(9) of the previous section.

\subsubsection{{\rm CO}$_2$ case}

\begin{figure}[!t]
\begin{center}
\includegraphics[width=0.99\columnwidth]{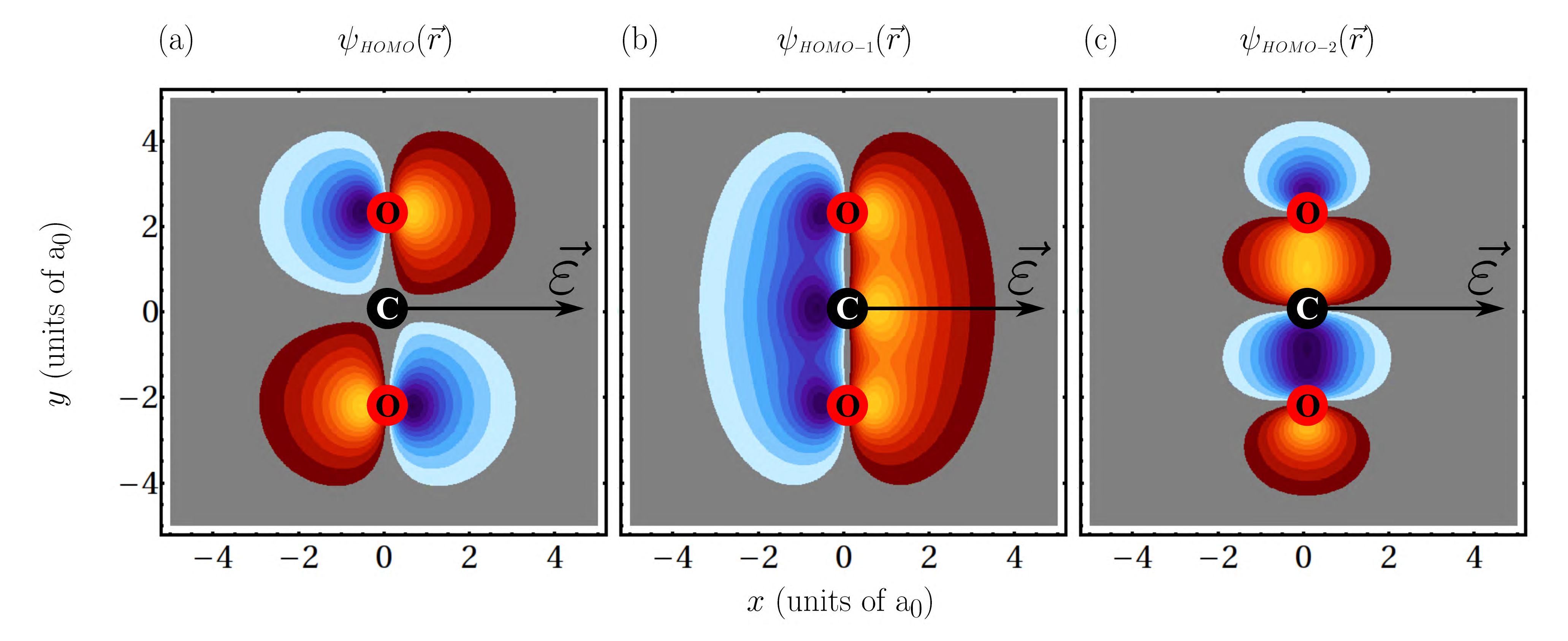}
\caption{\label{fig5}(Color online) Two-dimensional representation of the (a) HOMO, (b) HOMO-1 and (c) HOMO-2 probability amplitude distributions of CO$_2$. The change of color denotes a change of sign of the orbital wave function.}
\end{center}
\end{figure}

\begin{figure}[!t]
\begin{center}
\includegraphics[width=0.99\columnwidth]{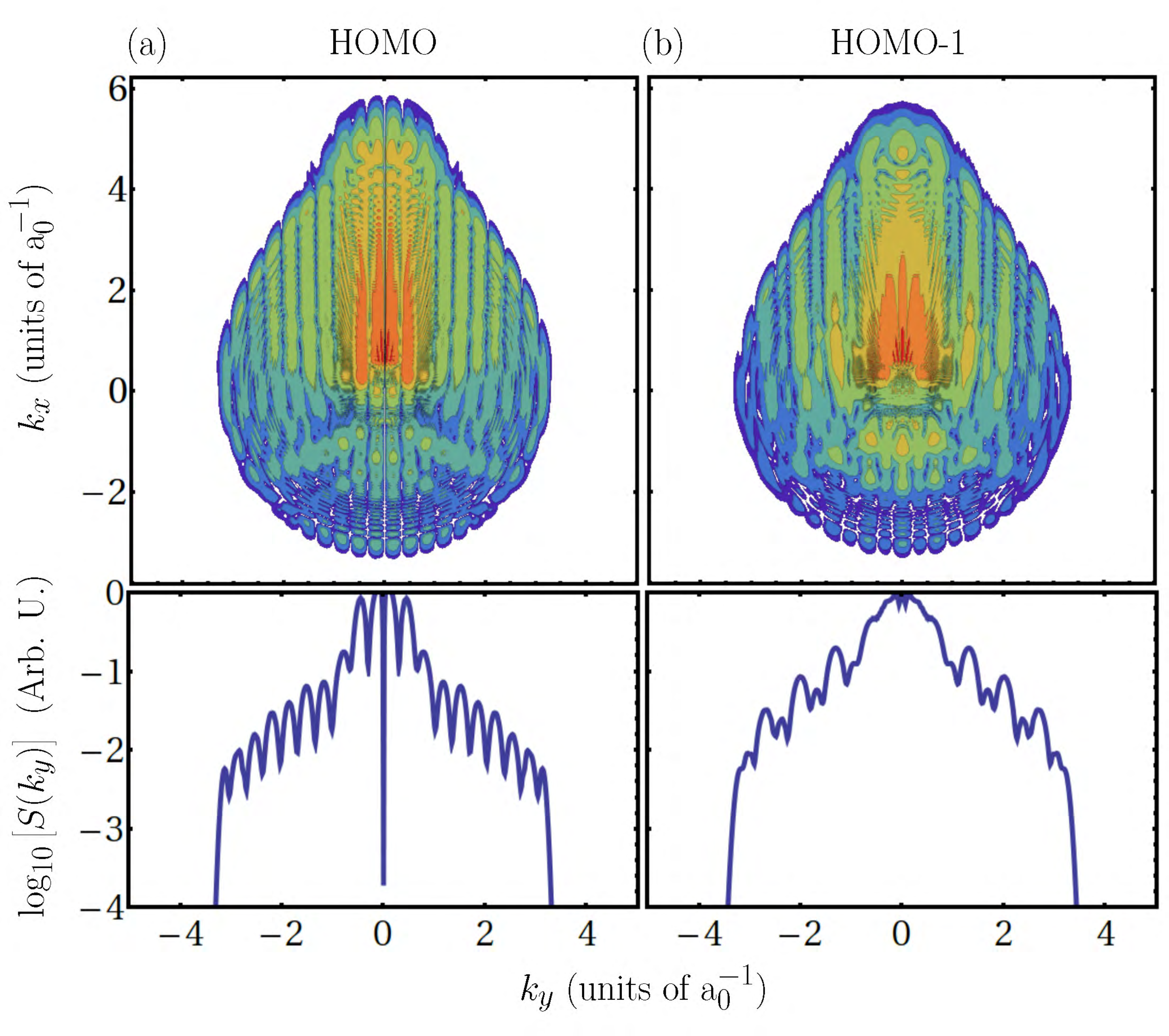}
\caption{\label{fig6}(Color online) Electron diffraction patterns of a symmetrically stretched CO$_2$ ($R=4.8$\,\AA) in the momentum space (upper panels) for the HOMO (a) and HOMO-1 (b) orbitals under the same excitation conditions as in Fig.\,\ref{fig4}. The lower panels display the averaged diffraction spectra $S(k_y)$ on a logarithmic scale normalized to the maximum value.}
\end{center}
\end{figure}

For CO$_2$, we considered ionization out of either the HOMO, or the HOMO-$n$ $(n=1,2)$.  These orbitals are sufficiently close in energy to each other in the field-free molecule to play an important role in the laser-driven ionization process for the choice of laser parameters under consideration \cite{Pavicic2007, Abu2009, Fowe2010}. Their well-known LCAO  structures have been recalled in the preceding section. They are shown in Fig.\,\ref{fig5}. The differences in symmetry and nodal properties between these three molecular orbitals (MO) are obvious from this figure.  The most important feature in this respect and for the purpose of the present discussion is   that the HOMO and HOMO-2 are anti-symmetric with respect to the mirror plane orthogonal to the molecular axis and containing the C atom, while the HOMO-1 is symmetric. This is at the origin of the different fringe structures found in both the 2D photoelectron momentum distribution and $k_x$-averaged spectra, \textit{i.e.} diffraction patterns $S(k_y)$, as shown in Fig.\,\ref{fig6}, lower panels.

The HOMO and the HOMO-2 (panels (a) and (c) of Fig.\,\ref{fig5}) are both of symmetry $b_2$ in $C_{2v}$, and have a general LCAO composition of the form
\begin{equation}\label{CO2_HOMO}
\Phi(\vec r) \propto f_{\rm O}(\vec r -\vec R) + f_{\rm C}^{b2}(\vec r) - [\mathcal{S} f_{\rm O}](\vec r +\vec R)
\end{equation}
where $f_{\rm O}(\vec r)$ will generally be a hybrid Oxygen atomic orbital, and $f_{\rm C}^{b2}(\vec r)$ denotes some function of symmetry $b_2$ centered on the C atom. $R$ here denotes the C-O internuclear distance. Similarly, one can write
\begin{equation}\label{CO2_HOMOt}
\tilde \Phi(\vec k,t) \propto e^{i\vec k.\vec R} \tilde f_{{\rm O},t}(\vec k) + \tilde f_{{\rm C},t}^{b2}(\vec k) - e^{-i\vec k.\vec R} [\mathcal{S}\tilde f_{{\rm O},t}](\vec k)
\end{equation}
such that, writing once again
\begin{equation}
f_{{\rm O},t}=f_{{\rm O},t}^a+f_{{\rm O},t}^b \qquad \mathrm{and} \qquad \mathcal{S}f_{{\rm O},t}=f_{{\rm O},t}^a-f_{{\rm O},t}^b
\end{equation}
the momentum wave function now reads
\begin{eqnarray}\label{CO2_HOMO_as}
\tilde \Phi(\vec k,t) & \propto & 2 \cos(k_yR)\tilde f_{{\rm O},t}^b(\vec k) + \tilde f_{{\rm C},t}^{b2}(\vec k) \nonumber\\
                      &         & + 2 i \sin(k_yR)\tilde f_{{\rm O},t}^a(\vec k)
\end{eqnarray}

\begin{figure}[!t]
\begin{center}
\includegraphics[width=0.99\columnwidth]{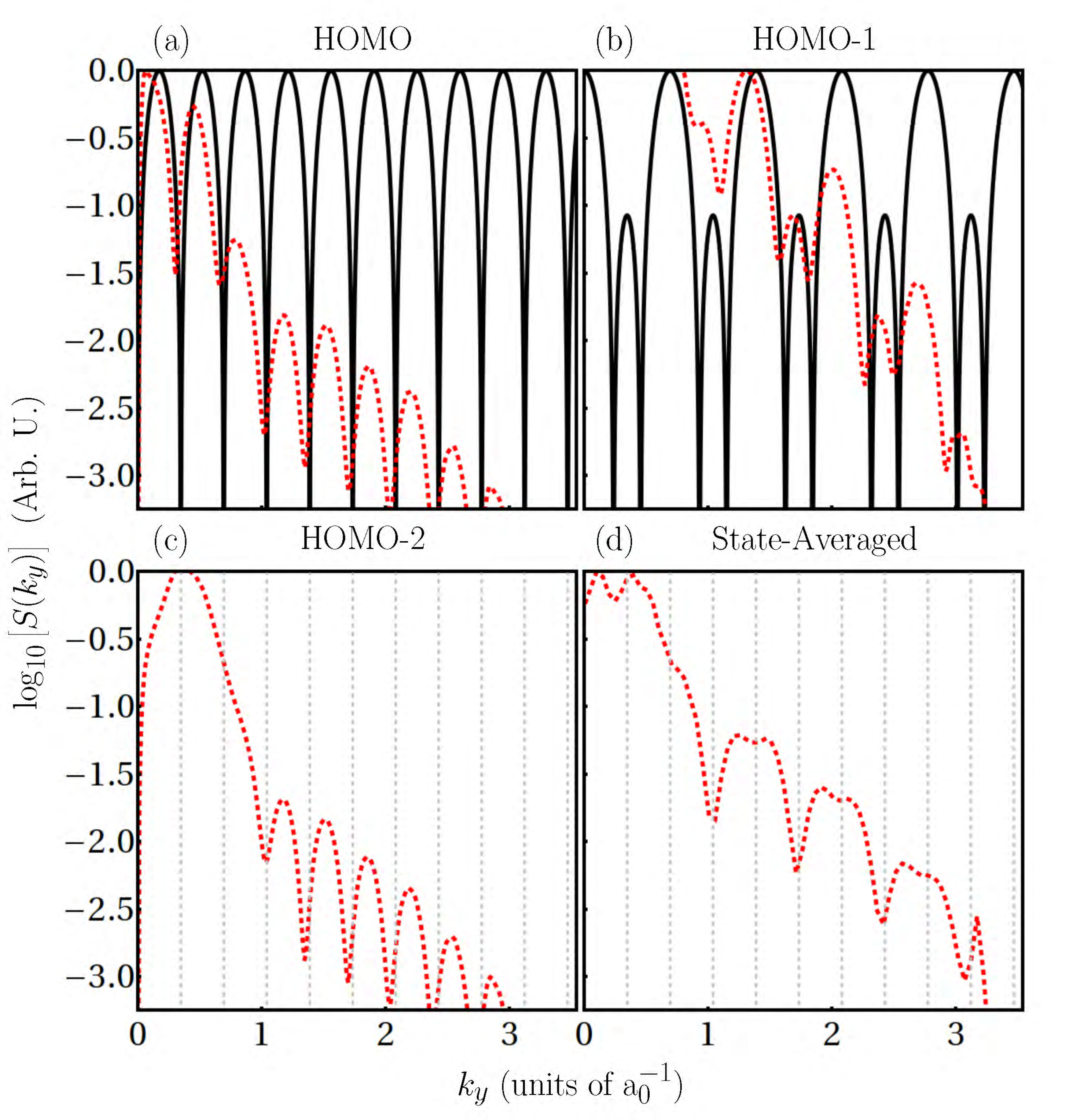}
\caption{\label{fig7}(Color online) Averaged diffraction patterns $S(k_y)$ on a logarithmic scale normalized to the maximum value calculated from the three different initial states of a symmetrically stretched CO$_2$ molecule with a C-O distance of $R=4.8$\,\AA\ under the same excitation conditions as in Fig.\,\ref{fig1}. Panels (a) to (c) correspond respectively to the HOMO, HOMO-1 and HOMO-2. Black solid lines show the theoretical reconstruction through the inversion algorithm and the red dotted lines show the numerical results. Panel (d) displays the same observable for an incoherent superposition of the three orbitals with equal weights. Gray dotted lines indicate the expected positions $k_y = n\pi/R$ of the minima of the function $\sin^2(k_yR)$.}
\end{center}
\end{figure}

As shown in Fig.\,\ref{fig7}(a) and (c), in the cases of the HOMO and HOMO-2, the diffraction patterns $S(k_y)$ have fringes  at large $k_y$ that are mainly governed by a $\sin^2(k_yR)$ term, with zeroes at $k_y = n\pi/R$. The curves in black (solid) lines represent this pure $\sin^2(k_yR)$ diffraction pattern. The calculated patterns shown in red (dashed) lines follow closely this behavior and exhibit zeroes at the predicted positions. The last term of Eq.(\ref{CO2_HOMO_as}) is thus seen to dominate the calculated spectrum in this range of $k_y$. At smaller values of $|k_y|$, the zeroes of this $\sin^2(k_yR)$ part are masked by the contributions of the first terms of Eq.(\ref{CO2_HOMO_as}), which die out rapidly as $|k_y|$ increases. Indeed, $\tilde f_{{\rm O},t}^b(\vec k)$ and $\tilde f_{{\rm C},t}^{b2}(\vec k)$ are essentially the Fourier transforms of the contributions arising from the diffuse $2p_y$ atomic orbitals centered on O and C, which are narrower in $k$-space than the contribution $\tilde f_{{\rm O},t}^a(\vec k)$ arising from the $2s$ atomic orbital.

The HOMO-1 is a $\pi_u$ bonding orbital and is of symmetry $a_1$ in $C_{2v}$. Its analysis thus proceeds as done above for the HOMO of N$_2$, except that here the LCAO pattern will be a 3-center   one. In fact, of the $n=2$ atomic orbitals of O and C, only the $2p_x$ orbital is simultaneously of $\pi$ symmetry in $D_{\infty h}$ and $a_1$ in $C_{2v}$. The HOMO-1 can thus be well described by the combination
\begin{equation}\label{CO2_HOMO-1}
\Phi_{\pi_u}(\vec r) \propto f_{\rm O}(\vec r -\vec R) + f_{\rm C}(\vec r) + f_{\rm O}(\vec r +\vec R)
\end{equation}
Assuming that this structure is conserved during the propagation in the presence of the field, the above reasoning leads to
\begin{equation}\label{CO2_HOMO-1_as}
\tilde \Phi_{\pi_u}(\vec k,t) \propto 2 \cos(k_yR)\tilde f_{{\rm O},t}(\vec k) +  \tilde f_{{\rm C},t}(\vec k)
\end{equation}
If we assume that $f_{\rm C}(\vec r) \simeq f_{\rm O}(\vec r) = g(\vec r)$, \textit{i.e.} that O and C have the same $2p_x$ orbitals, we get
\begin{equation}
\Phi_{\pi_u}(x, y, t) \simeq g_t(x, y-R) + g_t(x, y) + g_t(x, y+R)
\label{homo-1}
\end{equation}
and the diffraction pattern $S(k_y)$ should exhibit a double peak structure just as $[1 + 2 \cos(k_yR)]^2$ with zeroes for
\begin{equation}
k_y = \left(n+\frac{1}{3}\right)\frac{2\pi}{R}\;\;\textrm{and}\;\;k_y = \left(n+\frac{2}{3}\right)\frac{2\pi}{R}
\end{equation}
and maxima at $k_y = n\pi/R$. The profile of the pure $[1 + 2 \cos(R\,k_y)]^2$ diffraction pattern is shown in the panel (b) of Fig.\,\ref{fig7} in black (solid) line with the numerical result shown in (red) dotted line. This comparison confirms again the validity of the proposed inversion algorithm.

A less stringent assumption would be that $f_{\rm C}(\vec r) = \gamma\,g(\vec r)$ while $f_{\rm O}(\vec r) = g(\vec r)$, \textit{i.e.}
\begin{equation}
\Phi_{\pi_u}(x, y, t) = g_t(x,y-R) + \gamma\,g_{t}(x,y) + g_{t}(x,y+R)
\label{lcao}
\end{equation}
and in which case
\begin{eqnarray}
|\tilde{\Phi}_{\pi_u}(k_x, k_y, t)|^2 & \propto & [ 2 + \gamma^2 +2 \cos(2 R k_y)\nonumber\\
                              &         &   +\,4\gamma \cos(Rk_y) ]\,|\tilde{g}_t(k_x,k_y)|^2
\end{eqnarray}
One can then show that the ratio, $\sigma$, of the amplitudes of two successive peaks in the diffraction pattern resulting from this is related to the ratio  $\gamma$ of the LCAO coefficients of the $2p_{x}$ atomic orbitals of C and O in this molecular orbital. Indeed, we obtain $\sigma \simeq (\gamma-2)^2/(\gamma+2)^2$ or, equivalently, $\gamma = (2-2\sqrt{\sigma})/(1+\sqrt{\sigma})$. The intensity ratio of the double peak structure of the calculated diffraction pattern  for the HOMO-1 orbital, shown in panel (b) of Fig.\,\ref{fig7} in red (dashed) line, is $\sigma \simeq 1/11.8$ which gives a parameter $\gamma \simeq 1.10$, in support of the simpler expansion of Eq.(\ref{homo-1}) and in agreement with the pictorial representation of the HOMO-1 orbital in Fig.\,\ref{fig5}(b).

\subsubsection{Generalization}

A number of general features of the above analysis, permitting the retrieval of the geometrical and orbital structure of the molecule from observable photoelectron momentum distributions, and constituting the LIED inversion algorithm, is worth noting here.

We first note that, at large values of $|k_y|$, $S(k_y)$ is dominated by a single oscillatory term, associated with that atomic component of the initial molecular orbital that is widest in the momentum space. Thus, disregarding  the modulation of the interference pattern in $S(k_y)$, in particular its decay  as $k_y$ increases,  we can concentrate on this oscillatory pattern at large $|k_y|$ to retrieve informations on the bond-length(s) of a symmetric linear molecule. More precisely the positions of the zeroes and  peaks of the calculated spectra are accurately predicted by a simple periodic function of $k_y R$, the central element of the inversion algorithm, so that one can recover the sought-for geometrical information   through  relations of the type $R =n \pi / \Delta k_y$ where $\Delta k_y$ denotes the $k_y$-spacing separating two consecutive maxima. This is the quantitative aspect of the inversion algorithm.

We have been able to reproduce, by using just the simple formula evoked above, the regular oscillation patterns found in the photoelectron parallel momentum spectrum, $S(k_y)$, in all cases and trace these patterns back to the symmetry and nodal, or rather LCAO structure of the initial molecular orbital: for example, the simple structure of the spectrum in the case of the HOMO [Fig.\,\ref{fig7}(a)] as opposed to that of the  HOMO-2 [Fig.\,\ref{fig7}(c)], reflects the simple LCAO structure of this orbital. The double peak structure of the \mbox{HOMO-1} [Fig.\,\ref{fig7}(b)] reflects also the 3-centered LCAO structure of this orbital. This qualitative aspect of the inversion algorithm distinguishes it from previous, existing LIED analysis approaches, and is the feature that depends most on the alignment condition as it is deeply rooted in the structure and symmetry of the molecular system.

We turn next to the  role and importance of the model assumptions, and assess  the robustness of this LIED analysis, as an imaging technique for the retrieval of geometrical information, with respect to the variations in the model parameters. This is conducted by examining separately those concerning molecular degrees of freedom and the ones in relation with the laser field.

\subsection{Robustness with respect to molecular degrees of freedom}
\label{sec4_TTND_B}

It is useful to make here a distinction  between internal and external molecular degrees of freedom when referring to the various assumptions made above. Assumptions on internal degrees of freedom are those made in describing electronic and vibrational dynamics. Assumptions on external degrees of freedom are those concerning the  rotational dynamics. Let us examine the consequences of these approximations successively.

\subsubsection{Internal degrees of freedom: electronic dynamics}

Within the framework of the SAE approximation, we have considered the ionization and the subsequent electronic dynamics to start from one of the highest occupied orbitals of the CO$_2$ molecule. The resulting spectra for the HOMO-$n$ ($n=0$, 1, 2)   gathered in Fig.\,\ref{fig7} show that: (\textit{i})  Very sharp minima at the momenta $k_y$ are predicted by the inversion algorithm from which the value of the C-O bond length $R$ can be obtained. More precisely, from the spectra shown in the figure, one obtains $R=4.92$\,\AA \ with an error of less than 3\% over 8 peaks, when compared with the input parameter $R = 4.80$\,\AA. (\textit{ii}) The observation of a periodic sequence of minima is clearest in the high energy region of the spectra, which incidentally corresponds mainly to backward scattering.

Without relaxing yet the SAE approximation, we have also shown in panel (d) of Fig.\,\ref{fig7} how an incoherent superposition of the three HOMOs with equal probability still results in a spectrum with sharply defined minima from which the geometrical information can be extracted with a comparable accuracy.

We now argue that, provided the  molecule is perfectly aligned as assumed, going beyond the SAE should presumably not affect the readability of the diffraction pattern. Indeed, the symmetry character of the quantum state holds for the $N$-electron, fully correlated, time-dependent wave function, as well as for the orbitals (one electron wave functions). Thus, all what was written above concerning the SAE wave function should also apply to the $N$-electron wave function.

Now, the photoelectron momentum distribution can certainly be analyzed in terms of properties of the \textit{one-electron} (time-dependent) density matrix, $\gamma_1(t)$, derived from the exact $N$-electron state $|\Psi_N(t)\rangle$, hence from its eigenfunctions, the natural orbitals, $\varphi_i(\vec r,t)$ \cite{natorb}.

For the purpose of this demonstration, let us first define the $N$-electron density matrix
\begin{equation}
\rho_N(t) \equiv |\Psi_N(t)\rangle\langle\Psi_N(t)|
\end{equation}

The \textit{one-electron} density matrix $\gamma_1(t)$, which can be obtained from the partial trace
\begin{equation}
\gamma_1(t) = N\,{\rm Tr}_{N\!-\!1}\,\rho_N(t)
\end{equation}
over the remaining $(N-1)$ electrons, defines the set of natural orbitals using
\begin{equation}
\gamma_1(t) = \sum_i n_i\,|\varphi_i(t)\rangle \langle\varphi_i(t)|\,. \label{1e_density_matrix}
\end{equation}
The exact time-dependent electron momentum density can then be written as
\begin{equation}\label{rho_k}
\tilde \rho_1(\vec k, t)= \sum_i n_i\,|\tilde \varphi_i(\vec k,t)|^2
\end{equation}
For symmetric linear molecules aligned perpendicular to the laser polarization, whose common reduced point group is $C_{2v}$, the natural orbitals can only be of $a_{1(2)}$ or $b _{1(2)}$ symmetry, and can always be expanded in a complete atomic basis. Their asymptotic momentum distribution should be either of the form given in Eq.(\ref{CO2_HOMO_as}), for a general $a_{1(2)}$ orbital, or of the form of Eq.(\ref{CO2_HOMO-1_as}), for a general $b_{1(2)}$ orbital. It follows then that the exact photoelectron momentum distribution should be the incoherent sum, defined by Eq.(\ref{rho_k}), of the natural orbitals' asymptotic densities of the form
\begin{align}\label{NO_sum_k_dist}
|\tilde \Phi(\vec k, t)|^2 \propto & \big|a \cos (\vec k.\vec R)\tilde f_{1} (\vec k) + b \sin (\vec k.\vec R)\tilde f_{2} (\vec k)\nonumber\\&  + c\tilde f_{3} (\vec k)\big|^2
\end{align}
where $\tilde f_{1(2,3)}$, are some functions of the free electron momentum $\vec k$ (here in 3D), whose explicit forms depend  on the natural orbitals.

The same type of periodic fringe pattern as shown in panel (d) of Fig.\,\ref{fig7}, with readable periodic zeroes and maxima at large momenta, should thus characterize this photoelectron momentum distribution. Work is in progress in our group in order to study in detail this process and the influence of electron correlations in the LIED process.

\begin{figure}[!t]
\begin{center}
\includegraphics[width=0.99\columnwidth]{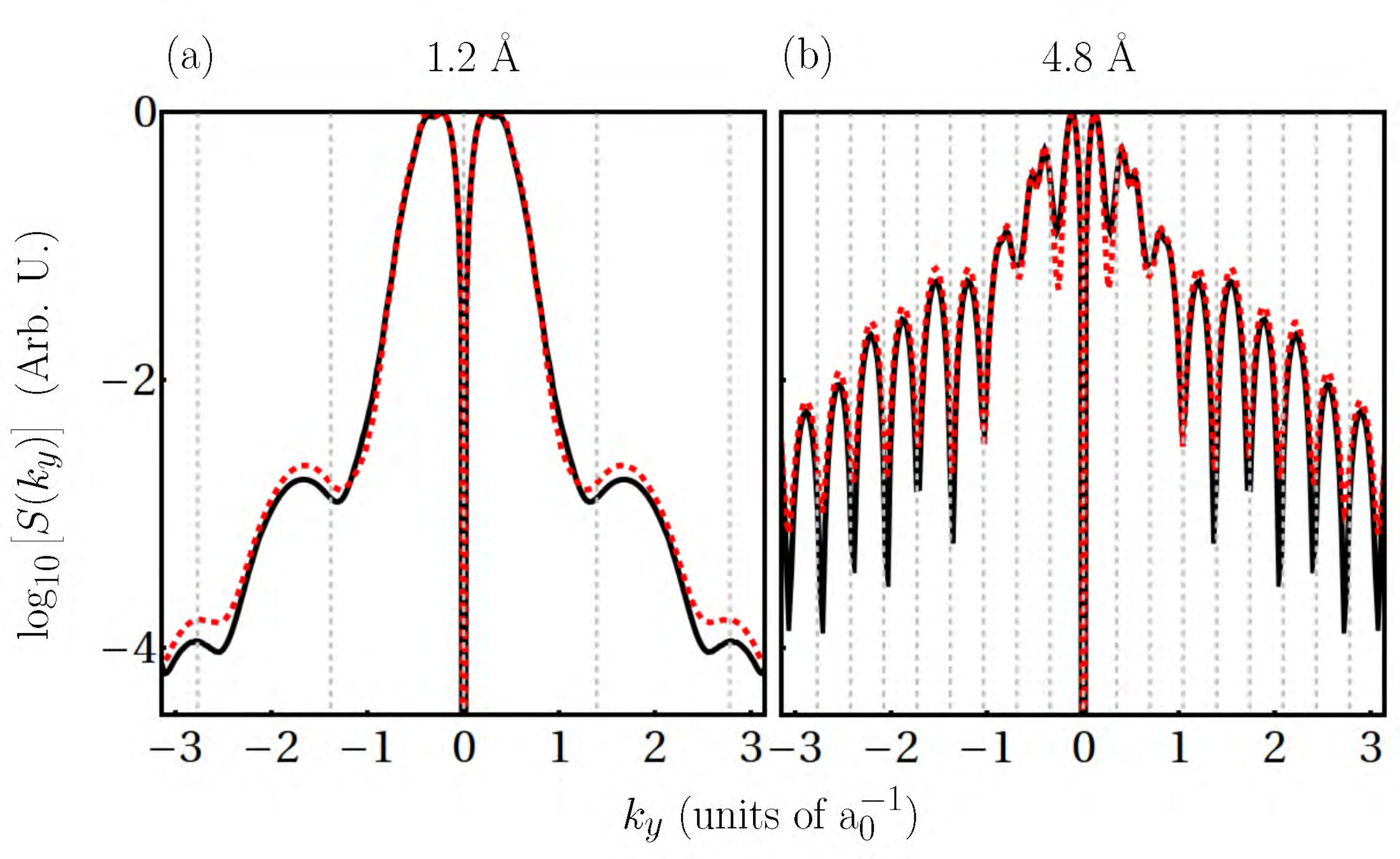}
\caption{\label{fig8}(Color online) High energy ($|k| > 3.15$\,a.u) averaged diffraction patterns $S(k_y)$ from the HOMO of CO$_2$ in its equilibrium ($R_0=1.2$\,\AA, panel a) and symmetrically stretched ($R_0=4.8$\,\AA, panel b) geometries, under the same excitation conditions as in Fig.\,\ref{fig1}. Black solid lines show the results in the absence of vibration (fixed $R$) and the red dotted lines take into account a vibrational dispersion of $\Delta R = \pm 1$\,\AA. The gray dotted lines indicate the expected positions of the minima of the function $\sin^2(R_0\,k_y)$, for the corresponding values of the internuclear distance. See text for details.}
\end{center}
\end{figure}

\subsubsection{Internal degrees of freedom: vibrational dynamics}

The structural reading made above of the diffraction patterns refers to a molecule in a fixed nuclear geometry configuration, perfectly aligned perpendicular to the linear polarization vector $\vec\varepsilon$. In reality, molecules are subject to vibrations and rotations, and these geometrical parameters are usually distributed over some range depending on the vibrational and rotational state in which it is initially prepared. The possible blurring due to molecular vibrations in the observed diffraction spectra can be taken into account through an incoherent average of the diffraction signal over independent calculations performed for different internuclear distances. Here we choose a range of $R$ going from $(R_0-1\textrm{\AA})$ to $(R_0+1\textrm{\AA})$ both with $R_0 = 1.2$\,\AA\ and 4.8\,\AA, under the same excitation conditions as in Fig.\,\ref{fig4}. The  weight associated with each value of the  internuclear distance $R$ is chosen as a Gaussian distribution ${\cal D}(R)$ normalized  over the finite interval $R \in [R_0-1\textrm{\AA}, R_0+1\textrm{\AA}]$
\begin{equation}
{\cal D}(R) = {\cal N} \exp \left(-\frac{[R-R_0]^2}{2\varsigma^2}\right)\label{distribution}
\end{equation}
where the standard deviation $\varsigma$, fixed at 0.2\,\AA, is much larger than that of the vibrational ground state associated with the symmetric stretch mode of CO$_2$. The incoherently averaged diffraction patterns are displayed in Fig.\,\ref{fig8}. They were obtained by retaining the higher energy signal only. We have chosen here $|k| > 3.15$\,a.u., which results in a significant increase of the contrast between constructive and destructive interference fringes. As can be seen from Fig.\,\ref{fig8}, the diffraction patterns are almost insensitive to the initial distribution of $R$. At both the equilibrium and  stretched geometries, the   contrast that remains after the vibrational motion has been  taken into account statistically in this way, allows for an accurate determination of the position of the fringes, which are not much affected by the bond length dispersion. A more quantitative estimation of this sensitivity can be reached by analyzing the Fourier transform of the asymptotic momentum amplitude associated with the HOMO, which reads
\begin{equation}
\tilde{\Phi}(\vec k,t) \simeq \sin( \vec k . \vec R  )\,\tilde{g}(\vec k),
\end{equation}
Replacing $\vec R$ by $\vec R_0 + \Delta\vec R$, setting $\vec k . (\vec R_0 + \Delta \vec R)= k_y.(R_0+ \Delta R)$, assuming perfect alignment, and considering the symmetric variations of both CO bond length, corresponding to a symmetric stretching mode, as done in the calculations shown in Fig.\,\ref{fig8},  one gets
\begin{equation}
\tilde{\Phi}(\vec k,t) \simeq [\sin(k_y R_0) + k_y \Delta R \cos(k_y R_0)]\,\tilde{g}(\vec k)
\end{equation}
A strong sensitivity with respect to $\Delta R$ therefore appears only for $|k_y| \gg 1/\Delta R$. In other words, for small bond length dispersions, only high energy fringes are significantly affected. This is confirmed in Fig.\,\ref{fig8}. The same analysis also applies to a non symmetric variation of the two C-O bond lengths, corresponding to a small amplitude vibrational motion in the anti-symmetric stretch mode. Likewise, a bending motion would give the same effect as the symmetric variations of the C-O bond lengths considered above. We thus expect diffraction patterns that are no less analyzable than the one shown in Fig.\,\ref{fig8}, and from which the C-O equilibrium bond length  can be extracted unambiguously.

\begin{figure}[!t]
\begin{center}
\includegraphics[width=0.99\columnwidth]{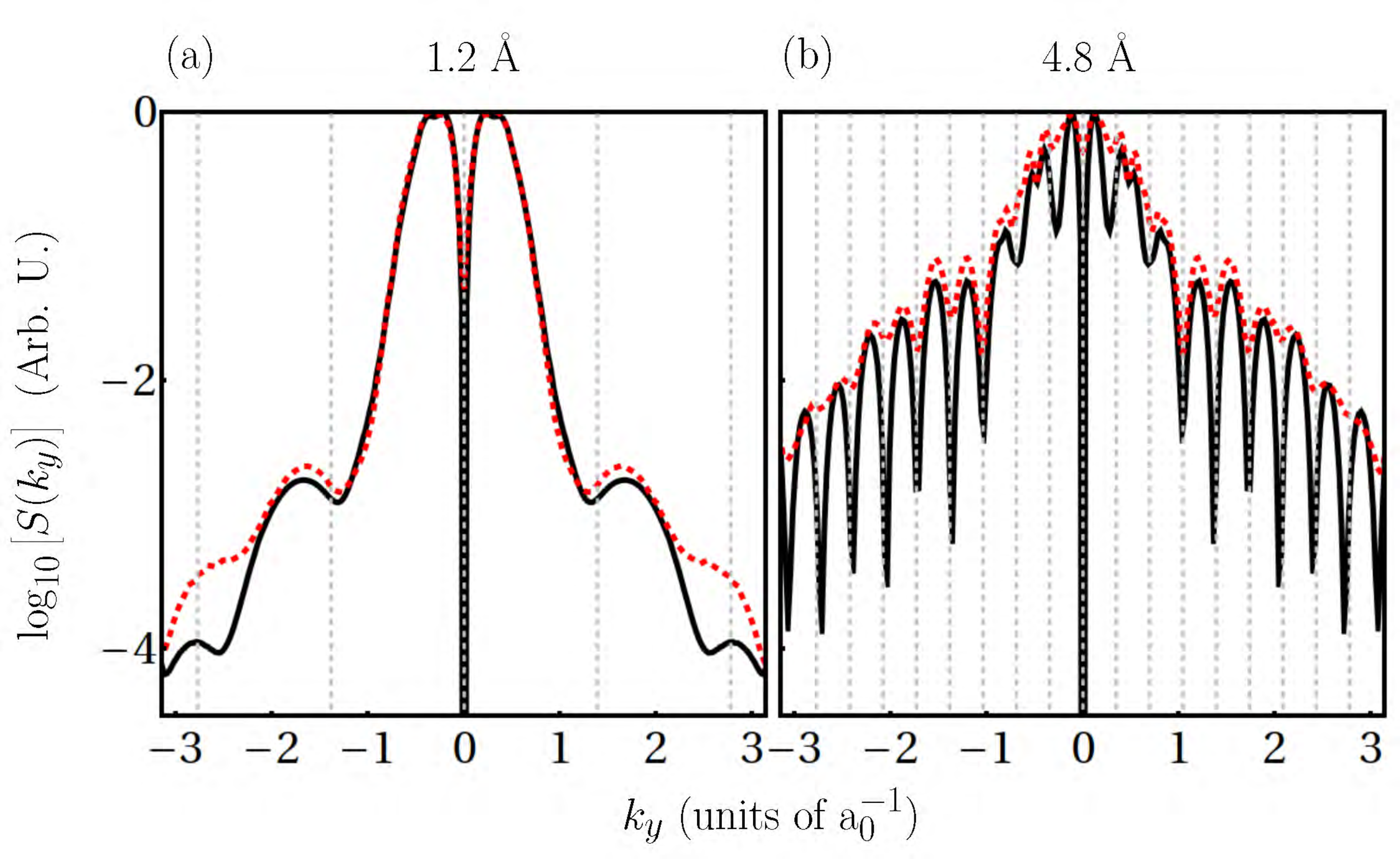}
\caption{\label{fig9}(Color online) High energy ($|k| > 3.15$ a.u) averaged diffraction patterns $S(k_y)$ from the HOMO of CO$_2$ in its equilibrium ($R=1.2$\,\AA, panel a) and symmetrically stretched ($R=4.8$\,\AA, panel b) geometries, under the same excitation conditions as in Fig.\,\ref{fig1}. The black solid lines show the results corresponding to perfect alignment at a fixed internuclear distance and the red dotted lines take into account an angular dispersion of 20$^{\circ}$ with the same internuclear distance. The gray dotted lines indicate the expected positions of the minima of the function $\sin^2(R\,k_y)$, for the corresponding values of the internuclear distance. See text for details.}
\end{center}
\end{figure}

\subsubsection{External degrees of freedom}

Finally, we examine the validity of the model assumptions with respect to the introduction of the external degree of freedom, that is the effect of angular dispersion. In that respect, it is useful to imagine that a first  laser pulse, an alignment pulse,  has   created a rotational wave packet which periodically re-phases, yielding thus, at the revival times, a strong   alignment of the CO$_2$ molecule perpendicular to the polarization direction of the subsequent ionizing laser pulse. As previously done for the vibrational motion, the effect of molecular rotation is simulated through similar statistical incoherent averaging over various molecular alignment angles $\theta$. The respective weight of each angle is determined by an analog of Eq.(\ref{distribution}), $D(\theta)$ now denoting a Gaussian angular distribution taken in the interval $\theta \in [0 , \pi]$. The standard deviation $\varsigma$ is chosen such that 90\% of the molecules are contained within a cone of aperture 20$^{\circ}$, corresponding to a strong alignment scenario that has recently been achieved experimentally for CO$_2$ \cite{CO2-align}. We apply to the scattered electron momentum spectrum a filter that retains only the high-energy electrons. The resulting signal, displayed in Fig.\,\ref{fig9}, is nicely contrasted and reproduces all the fringes of the aligned molecule. The only caveat is the weakness of the measured signal where the fringes are contrasted, but this could be overcome by a high repetition data recording. A quantitative analysis of the sensitivity with respect to $\Delta\theta$ leads to
\begin{equation}
\vec k . \vec R  = \left[1+\frac{k_x}{k_y}\tan(\Delta\theta)\right]\,k_y R \cos(\Delta\theta)
\end{equation}
which, for small $\Delta\theta$, results into:
\begin{equation}
\tilde{\Phi}(\vec k, t) \simeq \sin[k_y R (1+(k_x/k_y)\Delta\theta)]\,\tilde{g}(\vec k)
\end{equation}
giving a higher sensitivity to $\Delta\theta$ for $|k_y/k_x| \ll 1$. The fringes are actually slightly less contrasted for small $k_y$ values in Fig.\,\ref{fig9}(b). But we still expect diffraction patterns that are as analyzable as the ones shown in Fig.\,\ref{fig9}, and from which the C-O equilibrium and stretched bond lengths can be again unambiguously extracted.

\subsection{External field parameters}

\begin{figure}[!t]
\begin{center}
\includegraphics[width=0.99\columnwidth]{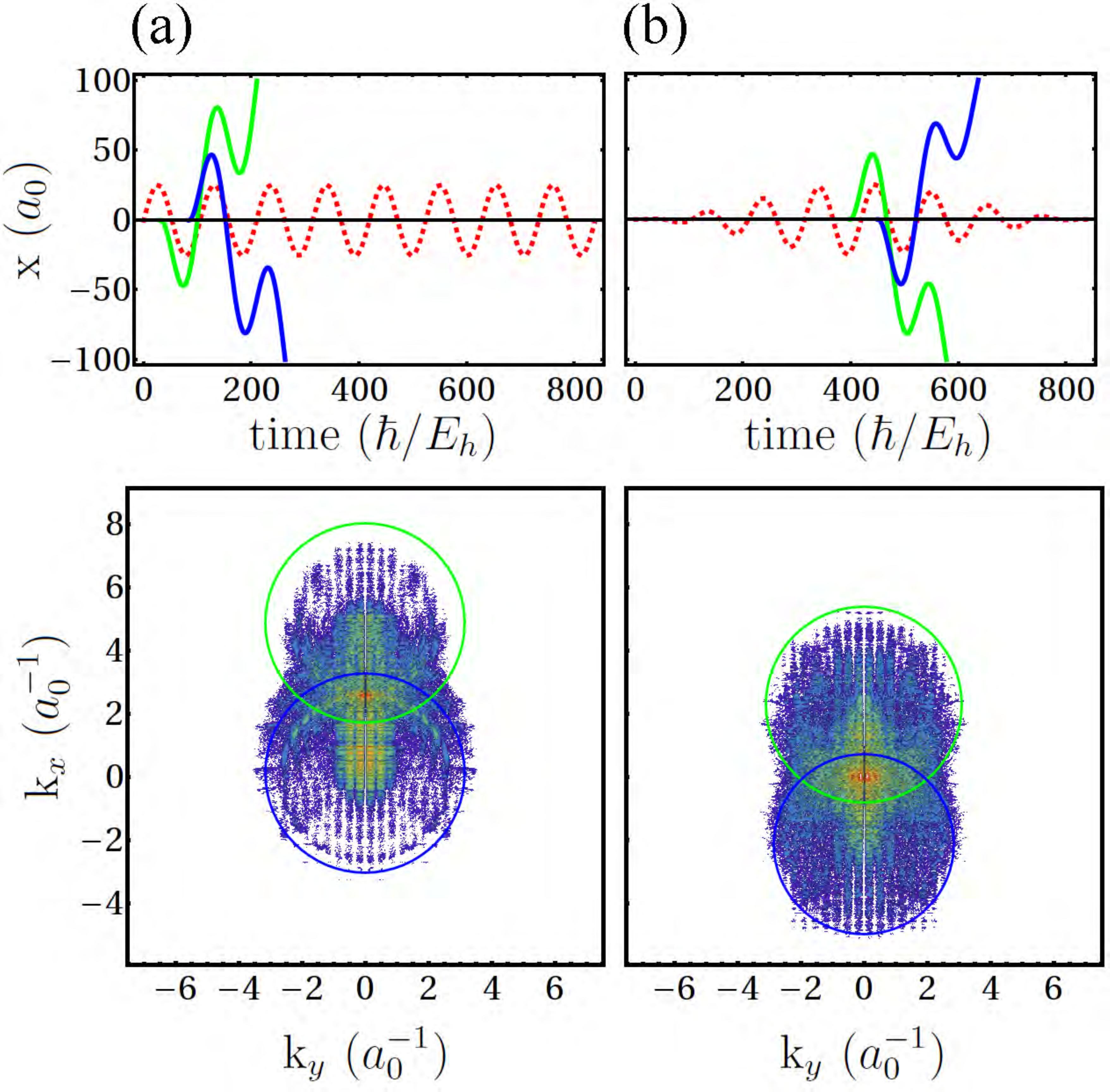}
\caption{\label{fig10}(Color online) Electron diffraction patterns of the HOMO of CO$_2$ with $R=4.8$\,\AA) in the momentum space representation (lower panels) for a rectangular pulse shape (panel a) and a sine-square pulse shape (panel b). The upper panels display the time variations in arbitrary units of the corresponding electric fields (in red dotted lines) and the two classical trajectories discussed in the text, one starting at positive electric field, in blue, and the other at negative electric field, in green.}
\end{center}
\end{figure}

In order to remove the limitations of an unrealistic single-cycle 800 nm laser pulse, we present results of calculations made with 8 optical cycles at the same carrier frequency. We will compare the results obtained with a rectangular pulse shape [upper row of Fig.\,\ref{fig10}(a)] with those obtained with a more realistic 10\,fs Full Width at Half Maximum (FWHM) sine-square pulsed excitation [upper row of Fig.\,\ref{fig10}(b)]. Both pulses cover about 8 optical cycles. The lower row of Fig.\,\ref{fig10} shows the diffraction pattern calculated for $R = 4.8$\,\AA\, assuming perfect alignment at fixed internuclear distance with this few-cycle excitation. With these long pulses, at least two electron wave packets are driven by the field in opposite directions before being both returned to the parent ion. Scattering signals arising from these different ``pathways'' interfere with each other, blurring out the diffraction pattern. Let us examine in some more details the role of the corresponding classical trajectories. The first row of Fig.\,\ref{fig10} depicts the time variation of the electric field, in red, and shows two classical trajectories corresponding to the maximum recollision energy, one (in blue) starting at a time at which the electric field is positive, and the other (in green) when the electric field is negative.

A recollision event arising from any of these trajectories occurs when the associated colored line intersects the $x = 0$ axis, where the incident (returning) electron momentum corresponds to a kinetic energy of $3.17\,U_p$. The electron associated with the first trajectory (blue line) scatters in every direction and continues to drift in the field, but since the recollision time is approximately $0.95\,T$, the residual shift in velocity is negligible (0.12\,a.u.). This gives rise to a signal mostly contained within a circle of radius $|k| \simeq 3.15$\,a.u. centered about $(k_x = 0.12$\,a.u., $k_y = 0)$. This is not the case for the second (green) trajectory where the electron continues to drift, after recollision, for some half integer multiple of the optical period $T$, giving rise to a large residual momentum drift of 4.87\,a.u. The associated signal is thus contained within a circle of same radius but centered about $(k_x = 4.87$\,a.u., $k_y = 0)$.

A different situation occurs with the sine-square pulsed laser excitation. If one performs the same calculation as described above for the maximal recollision energies starting at the times associated with the opposite maxima of the electric field, one finds slightly different recollision energies for the ``blue'' and ``green'' trajectories, respectively $2.8\,U_p$ and $3.1\,U_p$. These are also characterized by different residual drift momentum, respectively -2.12\,a.u. and 2.29\,a.u. The recollision energy is therefore reduced compared to the one obtained with a rectangular pulse shape for two reasons: (\textit{i}) the electric field is smaller in amplitude yielding different values of $U_p$, and (\textit{ii}) the electric field driving the electron back to the ion is constantly decreasing in amplitude, as opposed to the case of a rectangular pulse shape.

\begin{figure}[!t]
\begin{center}
\includegraphics[width=0.99\columnwidth]{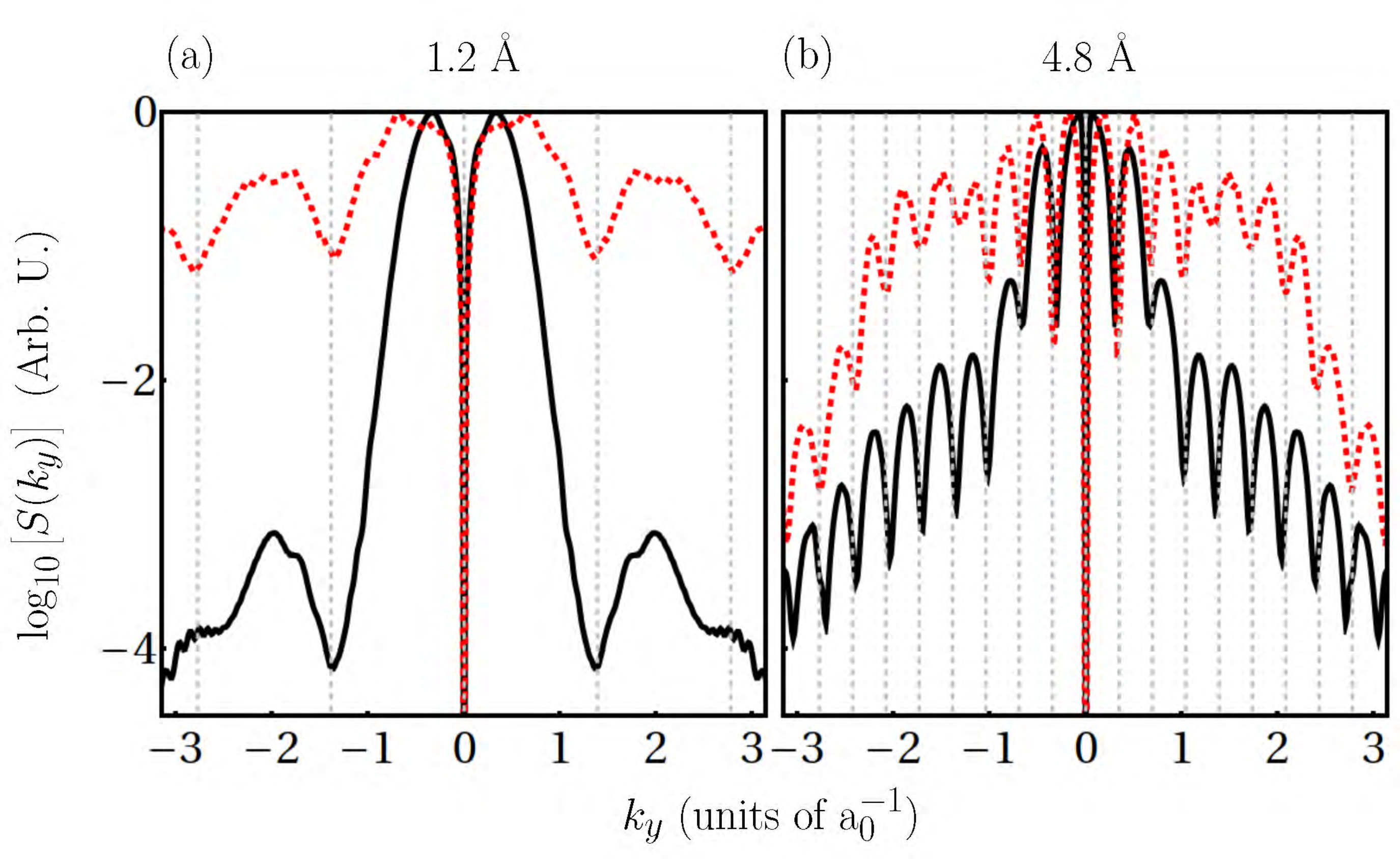}
\caption{\label{fig11}(Color online) High energy ($|k_x| > 3.15$\,a.u.) averaged diffraction patterns $S(k_y)$ from the HOMO of CO$_2$ in its equilibrium ($R=1.2$\,\AA, panel a) and symmetrically stretched ($R=4.8$\,\AA, panel b) geometries, under the same multi-cycle excitation conditions as in Fig.\,\ref{fig10}. Black solid lines show the results corresponding to a single-cycle excitation. The gray dotted lines indicate the expected positions of the minima of the function $\sin^2(k_y R)$, for the corresponding values of the internuclear distance.}
\end{center}
\end{figure}

With these long laser pulses, interferences between the different electron wave packets emitted in the forward and backward directions decrease the contrast obtained in the fringe spectrum. This difficulty can however be circumvented by applying a filter which restricts the $S(k_y)$ signal to high-energy electrons as shown in Fig.\,\ref{fig11} displaying this distributions in the high energy domain, with an average [Eq.(\ref{eq:average})] performed over $|k_x| > 3.15$\,a.u. only. Fig.\,\ref{fig11} shows the averaged diffraction patterns obtained for $R = 1.2$\,\AA\ in panel (a), and for $R = 4.8$\,\AA\ in panel (b), with a sine-square pulsed excitation of FWHM 10\,fs, represented in red dotted lines. In comparison, the black (solid) lines display the similar result for a single-cycle excitation, retaining again only high energy electrons to enhance the contrast. Once again, the contrast is perfectly sufficient to extract the corresponding C-O bond lengths.

\section{Conclusion}
\label{sec5}

In this paper, we have presented a detailed theoretical analysis of the ionization dynamics of the N$_2$ and CO$_2$ molecules in intense ultra-short linearly polarized laser fields at 800\,nm. Using two-dimensional single active electron effective potentials, we have calculated the momentum distribution of the emitted photo-electrons by solving the time-dependent Schr\"odinger equation for the electronic motion.

We have shown that, if the molecule is initially aligned perpendicular to the field polarization, a simple averaging and an inversion procedure can be used to determine the molecular bond length with an accuracy of a few percents. We have also shown that the photo-electron momentum distribution carries information on the structure and symmetry of the ionized molecular orbital. The robustness of the structure determination with respect to vibrational and rotational motions has also been demonstrated, and the inaccuracies introduced by such perturbations have been assessed and interpreted. These results throw some light on how to image such geometric and orbital information for a linear polyatomic molecule on an attosecond time‐scale by laser induced electron diffraction. Finally, taking into account the very short time acquisition of these molecular images, we also claim that a stroboscopical animation of the vibrational motion is achievable using the LIED strategy.

\section{Acknowledgements}

The authors would also like to acknowledge useful and stimulating discussions with Dr. Christian Cornaggia of the CEA IRAMIS in Saclay, France. M.P. and T.T.N.D. acknowledge the Natural Sciences and Engineering Research Council of Canada (NSERC) for financial supports. The authors also acknowledge supports from CFQCU (contract number 2010-19), from ANR (contracts ImageFemto ANR-07-BLAN-0162 and Attowave ANR-09-BLAN-0031-01), from FCS Digiteo - Triangle de la Physique (project 2010-078T - High Rep Image), and from the EU (Project ITN-2010-264951, CORINF), and the joint NSF (USA) - ANR (France) FRAMOLSENT project.

\bibliographystyle{apsrev4-1}

\end{document}